\documentclass[preprint,12pt]{elsarticle}
\usepackage{hyperref}
\usepackage{bm}
\usepackage{amsmath}
\usepackage{amsfonts}
\usepackage{subfig}
\usepackage{comment}

\journal{Journal of Space Safety Engineering}

\begin{document}

\begin{frontmatter}

\title{Assessing Collision Probability in Low-Thrust Deorbit}


\author{Shuta Fukii\fnref{myfootnote1}, Daisuke Sakai, Yasuhiro Yoshimura, \\Yuri~Matsushita, Toshiya Hanada, Yuki Itaya and Tadanori Fukushima}

\address{Kyushu University, 744 Motooka, Nishi-ku, Fukuoka, 819-0395, Japan\\ SKY Perfect JSAT Corporation, 8-1 Akasaka 1-chome, Minato-ku, Tokyo 107-0052, Japan}

\fntext[myfootnote1]{fukii.shuta.223@s.kyushu-u.ac.jp}



     
\begin{abstract}
End-of-life support of satellites is necessary to improve post-mission-disposal compliance rates for maintaining space environment. Deorbit mission with low thrust, e.g. a laser, induces a low-level deceleration on the target object that gradually lowers the target altitude. Since such a low-thrust trajectory is time-consuming, the risk of collision greatly influences the mission success rate. In this context, this paper assesses the collision risk during deorbit trajectories with low thrust. Furthermore, parametric studies for the relationship between the re-entry time and the risk of collision are performed.

\end{abstract}

\end{frontmatter}


\section{Introduction}
The number of objects orbiting around the Earth gradually increases, and this trend will put sustainable and safe space utilization at risk. 
In recent years, many commercial entities are considering the launch and deployment of large constellations of satellites, leading to a rapid increase in the number of satellites in the Low Earth Orbit (LEO) region. 
It is reported that constellation satellites more than 57000 will be planed to launch in next 10 years, and these satellites are deployed in LEO at an altitude of 300 km to 1000 km~\cite{alfano2020leo}~\cite{oltrogge2016collision}. It is observed that orbital congestion caused by a large number of satellites in LEO region can increase the likelihood of a collision event.
To curb the rapid rise in the number of objects in the LEO region, Active Debris Removal (ADR) and Post Mission Disposal (PMD) of satellites are required. 
Kawamoto et al.~\cite{KAWAMOTO2020178} indicated that these remediation options are effective if implemented properly.
Another commonly adopted mitigation measure is the so-called 25 years rule. This rule limits the post-mission orbital lifetime of satellites to be less than 25 years. 
However, this constraint may be insufficient once the large constellations have been deployed~\cite{virgili2016risk}.    

Radtke et al.~\cite{radtke2017interactions} investigated the interactions of space debris environment with the OneWeb constellation. This research conducted a risk assessment of OneWeb satellites for their entire operation and their influences on the orbital environment. 
Le May et al.~\cite{le2018space} evaluated the collision risk of large constellation satellites when the best or worst implementation case of mitigation guidelines. 

As shown in these studies, the high success rate of PMD is essential to maintain the orbital environment. Bastida-Virgili et al.~\cite{virgili2016risk} revealed that the successful implementation of 90$\%$ PMD on average is needed. 
On the other hand, other studies also suggest that even this requirement is insufficient~\cite{anz2018orbital}.
Since this requirement is substantially higher than today's moderate PMD success rate (approximately 60$\%$~\cite{lemmens2019esa}), it is unclear whether this guideline can be fully complied with.

Satellites in the LEO region use deorbit devices such as electric propulsion systems or drag sails to decrease their velocity and lower their altitude~\cite{SATO2020813}.
On the other hand, derelict or non-maneuverable satellites will require some external forces to complete the disposal. 
Thus, the End-Of-Life (EOL) deorbit of malfunctioned satellites must be assisted by on-orbit servicing missions. Various types of EOL support missions have already been proposed, including the use of electrodynamic tethers, lasers, and direct capture~\cite{RHATIGAN2020340}~\cite{phipps1996orion}~\cite{bischof2003roger}. 
This paper deals with the EOL support mission using laser-ablation-induced thrust. 
Generating thrust with laser ablation is also known as a suitable method for EOL support missions. 
The laser-ablation deorbit method has the advantage that the target satellite itself can be used as a propellant without physical contact, which makes the mission cost-effective and safe~\cite{tsuno2020impulse}. 

Satellites in the LEO region that have completed their mission phase usually lower their altitude during their disposal phase. Unless the satellites descend in a safe manner, they may potentially  generate many fragments if they were to collide with other  operational satellites or space debris. 
Thus, not only the operational phase but also the disposal phase should be carefully designed. 
Although some researches assessed the influence on the LEO environment caused by the deployment of large constellation satellites, the collision risk during a disposal phase is still unclear.

In this context, this paper assesses the collision risk of deorbiting with low thrust and focuses on the relation between the deorbit time and the collision risk. Designing deorbit trajectories that minimize fuel consumption~\cite{sims1999preliminary}~\cite{baldwin2012optimal} is outside the scope of this paper.
First, orbital propagation at different thrusting ranges is performed to obtain deorbit trajectories.
Then, the risk assessment of each orbital decay scenario is conducted using ESA MASTER--2009. The collision risk is evaluated on the basis of its collision probability and expected number of fragments. 
In order to reveal the characteristic of each scenario, the distribution of collision impact angles is also investigated.
Finally, this paper concludes that when designing deorbit mission, not only the efficiency of deorbit but also the collision risk that such a deorbit profile might introduce should be carefully examined.
It is noted that although this paper assumes the service satellite provides external forces by a laser for EOL support mission, the results presented in this paper can be applied to other deorbit missions using low-thrust.

\section{Preliminaries}

\subsection{Orbital propagation} 
This section obtains orbit trajectories of the target satellite as a preliminary step for risk assessment. Analytical methods based on general perturbation theory are often used to simulate orbital propagation where thrust forces are considered dominant. Although analytical propagation is computationally efficient, this study uses a numerical solution based on special perturbation theory because the acceleration by laser ablation is frequently switched on and off. To this end, Cowell's method is used in this paper, and the equations of motion in a Cartesian coordinate system is written as
\begin{equation}
    \ddot{\bm{r}} =  \frac{\mu}{|\bm{r}|^{3}}\bm{r}+\bm{F}+\bm{a}_{t} \label{eq:eom}
\end{equation}
where $\bm{r}$ is the position vector of the target satellite with respect to the Earth, $\mu$ is its gravitational parameter, and $\bm{F}$ is perturbing accelerations. In this paper, the effects of the Earth's geopotential, attraction due to the Sun and the Moon, atmospheric drag, and solar radiation pressure in the presence of Earth and lunar shadowing are considered perturbation forces. The acceleration generated by the laser ablation $\bm{a}_{t}$ is described as
\begin{equation}
    \bm{a}_t =  -\frac{F_t}{m_{\rm tar}}\frac{\bm{v}}{|\bm{v}|}
\end{equation}
where $F_t$ is the magnitude of thrust force generated by the laser ablation, $m_{\rm tar}$ is the mass of the target satellite, and $\bm{v}$ is the velocity vector of the target satellite relative to the laser satellite. Note that in this paper, the target satellite is the deorbiting spacecraft. 

This study adopts a box-wing type satellite as a target, which has the main cuboid body and two flat solar array paddles. This study assumes that a target satellite loses autonomous control and keeps rotating randomly (i.e., tumbling). According to \cite{2004orbital}, the projected area of a rectangular object in a random attitude can be calculated as a quarter of the surface area. Assuming that the target is a set of rectangles, the effective cross-sectional area is calculated as a quarter of the total surface area including solar panels.

\subsection{Deorbit scenarios}
\begin{figure}[tb]
    \centering
    \includegraphics[scale = 0.4]{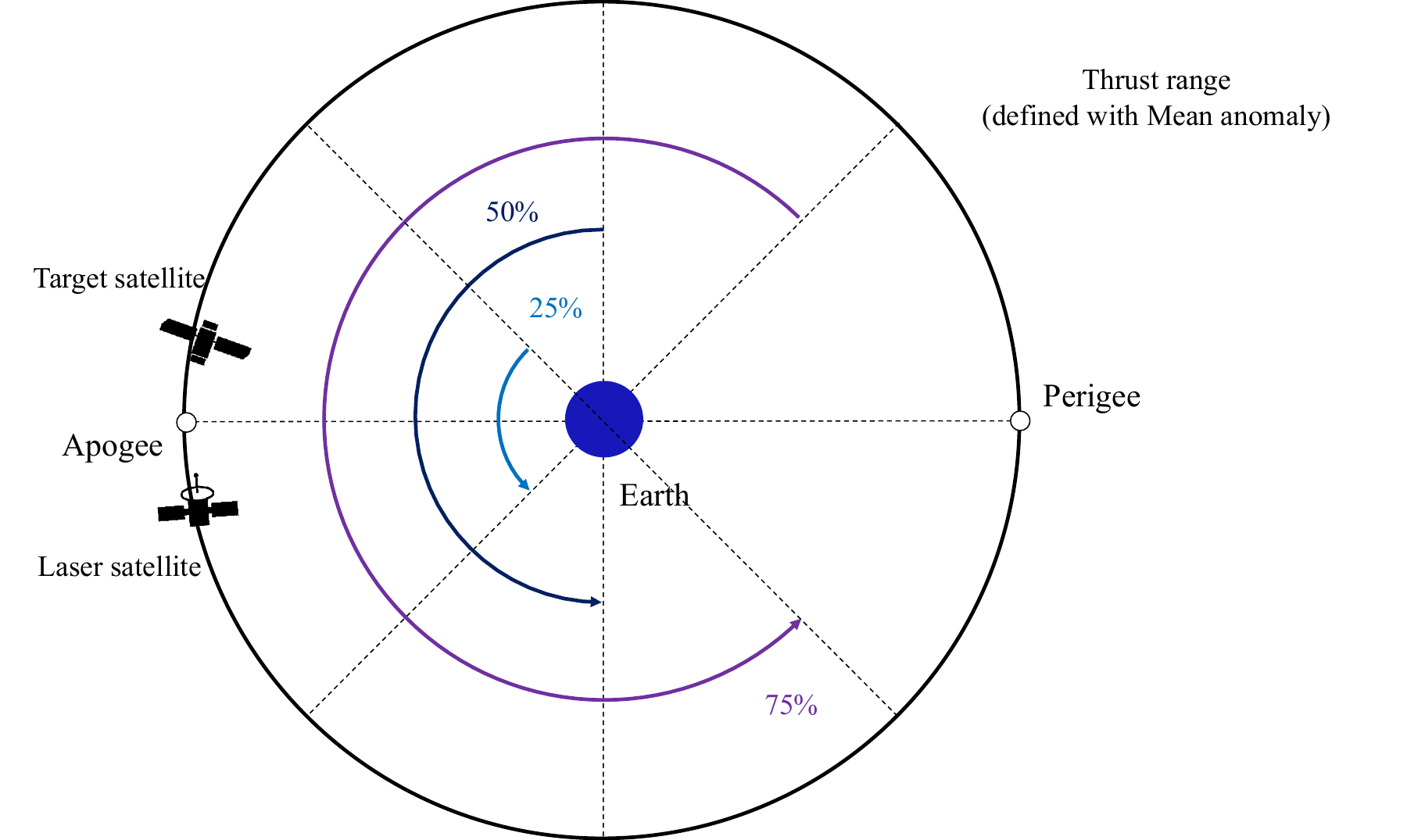}
    \caption{Thrust profiles. }
    \label{fig:thrust_profile}
\end{figure}

It is known that applying thrust around the apogee in the anti-velocity direction during the deorbit phase can effectively lower perigee altitude. 
This is because the closer the orbit is to the surface of the Earth, the larger the atmospheric drag is, which significantly decelerates the spacecraft. In this context, this study considers four scenarios of low-thrust profiles applied in the vicinity of apogee with low-thrust active 25$\%$, 50$\%$, 75$\%$, and 100$\%$ of the orbital period centered about apogee.
Geometrical illustration of the thrust profiles is described in Fig.~\ref{fig:thrust_profile}.
These thrusting scenarios are set using the mean anomaly $M$ as
\begin{equation}
    M = n(t-T)
\end{equation}
where $n$ is the mean motion and $T$ is the period of the orbit. The mean motion is the mean angular rate of the orbital motion. Therefore, products of percentages for each scenario and orbital period means thrusting period. For this study, we assumed that the laser satellite is in front of the target on the same orbit and the relative position between them is ideally maintained at a constant separation.  

\subsection{Collision analysis}
The collision flux analysis in this study is conducted using ESA's MASTER--2009. 
The mean number of collision $N$ can be calculated as 
\begin{equation}
\label{N_def}
    N = f A_c \Delta t
\end{equation}
where $f$ is the flux obtained from MASTER--2009 and $\Delta t$ is the duration. In this study, the collision cross section $A_c$ is calculated by assuming that the satellite is an ideal sphere as in the approach in \cite{le2018space}.
\begin{equation}
    A_c = \pi (r_{\rm tar} +r_{\rm deb})^2
\end{equation}
where $r_{\rm tar}$ and $r_{\rm deb}$ are the radii of the target satellite and debris objects, respectively. 
Again, in this paper, the target satellite is the deorbiting spacecraft and debris are debris fragment or other spacecraft.
Here we assume that the radius of the target satellite calculated as the radius of the sphere that has the same surface area as the target satellite. The radius of the target satellite $r_{\rm tar}$ is obtained via 
\begin{equation}
    r_{\mathrm{tar}} = \sqrt{\frac{S_{\mathrm{tar}}}{4\pi}}
\end{equation}
where $S_{\mathrm{tar}}$ is the surface area of the target satellite. 
Assuming that the collision number $X$ follows the Poisson distribution with parameter $N$, the probability for $k$ impacts is written as
\begin{equation}
    P(X=k) = \frac{N^k}{k!}e^{-N}
\end{equation}
Thus, the probability of at least one collision is 
\begin{equation}
    P(X\geq1) = 1 - e^{-N}
\end{equation}
The complete calculation and detailed deviation of the probabilities can be found in \cite{Klinkrand}.

\subsection{Collision classification}
In addition to analysing the risk for at least one collision, the catastrophic collision risk is also investigated.
A catastrophic collision is defined as a collision in which the satellites experience complete catastrophic fragmentation. In order to evaluate the damage of collision, the energy-to-mass ratio $\varepsilon$ is defined as
\begin{equation}
     \varepsilon = \frac{m_{\rm deb}v_{\rm rel}^2}{2 m_{\rm tar}}
\end{equation}
where $m_{\rm tar}$ and $m_{\rm deb}$ are the masses of the target satellite and the debris object, respectively, and $v_{\rm rel}$ is the relative velocity between the two objects. The collision that $\varepsilon$ exceeds 40 kJ/kg is considered catastrophic.

\subsection{Possible environmental effects}
For the purpose of assessing the possible environmental effects during the decay, the expected number of fragments are calculated on the basis of NASA's breakup model(\cite{johnson2001nasa}). The number of fragments $n_{\rm frag}$ larger than a given characteristic length $L_c$ is calculated as
\begin{equation}
\label{N_fra}
    n_{\rm frag} = 0.1 m^{0.75}L_c^{-1.71}
\end{equation}
where
\begin{align*}
    m &= m_{\rm tar} + m_{\rm deb} \qquad &{\rm if} \quad  \varepsilon \geq 40 \\
    m &= m_{\rm deb}v_{\rm rel}^2 \qquad &{\rm if}  \quad  \varepsilon < 40
\end{align*}

This study sets $L_c = 1$ mm, which means the fragments larger than 1 mm are considered.
By multiplying the mean number of collision in Eq.~\eqref{N_def} by the number of fragments in Eq.~\eqref{N_fra}, the expected number of fragments  $\mathbb{E}_i$ that can be generated at $i$-th collision event is 
\begin{equation}
    \mathbb{E}_i  =  n_{\mathrm{frag},i} N_i
\end{equation}
and adding them together for all possible events, the expected number of  fragments $\mathbb{E}_{\rm tot}$ is calculated. 
\begin{equation}\label{Exp_def}
    \mathbb{E}_{\rm{tot}} = \sum_{i} \mathbb{E}_i = \sum_{i} n_{\mathrm{frag},i} N_i
\end{equation}

\section{Numerical examples}

\subsection{Deorbit trajectory}\label{subsection:O-P}
Numerical simulations of orbital propagation are performed to obtain deorbit trajectories. Table~\ref{tab:Simulation conditions} shows the target's configuration and thrusting conditions. The mass is assumed to be 150 kg. The main cuboid body is assumed to be 1.3 m by 1.0 m by 1.0 m, whereas each solar array paddle is assumed to be 1.0 m by 1.3 m. The main body and the solar array paddles are 1.0 m apart. 
Table~\ref{tab:initial orbit} describes the initial orbital elements of the target satellite. The available period of low-thrust is assumed to be two years in total, and after that period, the orbit decays without thrust until deorbit.

\begin{table}[tb]
\centering
\caption{Simulation conditions.}
\begin{tabular}{cc}\hline\hline
Satellite mass, $m_{\rm tar}$ [kg] & 150 \\
Area to mass [m$^2$/ kg] & 0.0206\\
Thrusting force generated by low thrust, $F_{t}$ [mN]& 0.72\\
Available period of low thrust [years] & 2\\ \hline\hline
\end{tabular}
\label{tab:Simulation conditions}
\end{table}

\begin{table}[tb]
\centering
\caption{Initial orbit of the target satellite.}
\begin{tabular}{cc}\hline\hline
Epoch & 2026/01/01\\
Semi-major axis [km] & 7578.135\\
Eccentricity [-] & $1.0\times$10$^{-7}$\\
Inclination [deg] & 87.9\\
Argument of perigee [deg] & 0\\
Right ascension of node [deg] & 0\\
Mean anomaly [deg] & 0\\ \hline\hline
\end{tabular}
\label{tab:initial orbit}
\end{table}

\begin{figure}[tb]
    \centering
    \includegraphics[scale = 0.5]{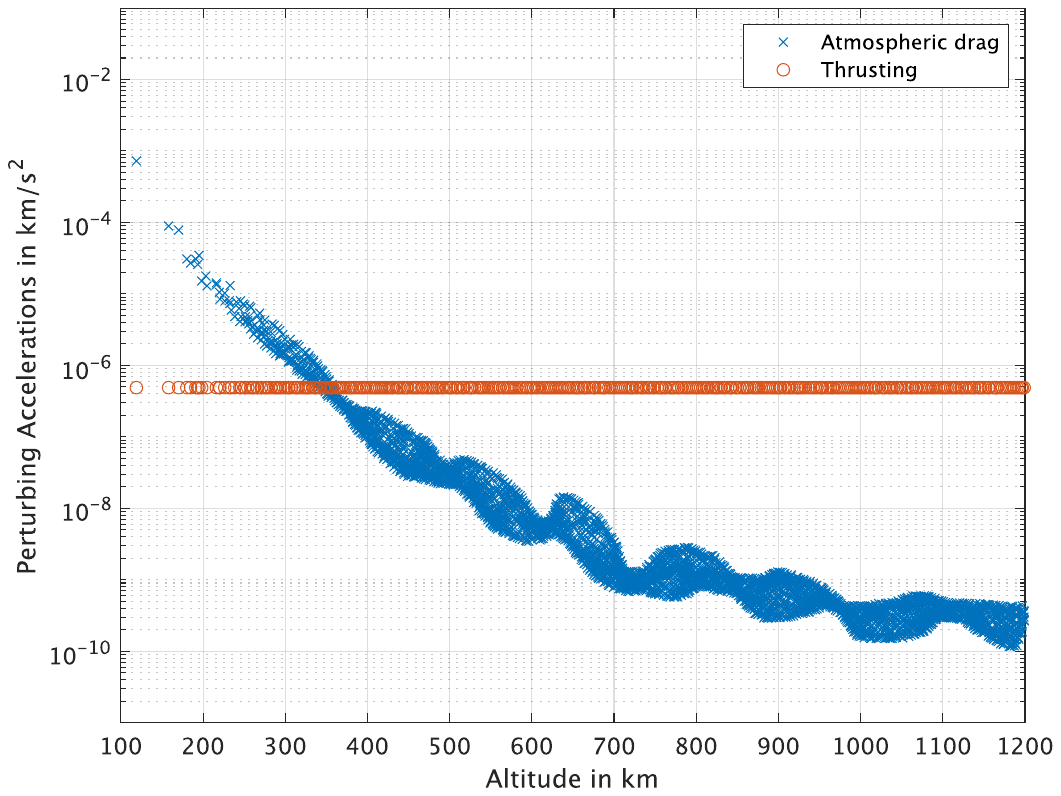}
    \caption{Comparison of the perturbing accelerations acting on deorbit. }
    \label{fig:perturbations}
\end{figure}

Figure~\ref{fig:perturbations} compares the magnitude of the perturbation accelerations due to the atmospheric drag and low thrust as a function of altitude. The magnitude of low thrust is  assumed to be constant at all altitudes and significantly influences the satellite's motion. Atmospheric drag is the perturbation force whose perturbation acceleration increases exponentially with decreasing altitude. After reaching 350 km or less at altitude, it produces a larger acceleration than the low thrust. This result confirms the effectiveness of initially reducing perigee altitude, yielding a large atmospheric drag when low thrust is applied only around the apogee.

\begin{figure}[tb]
    \centering
    \includegraphics[scale = 0.6]{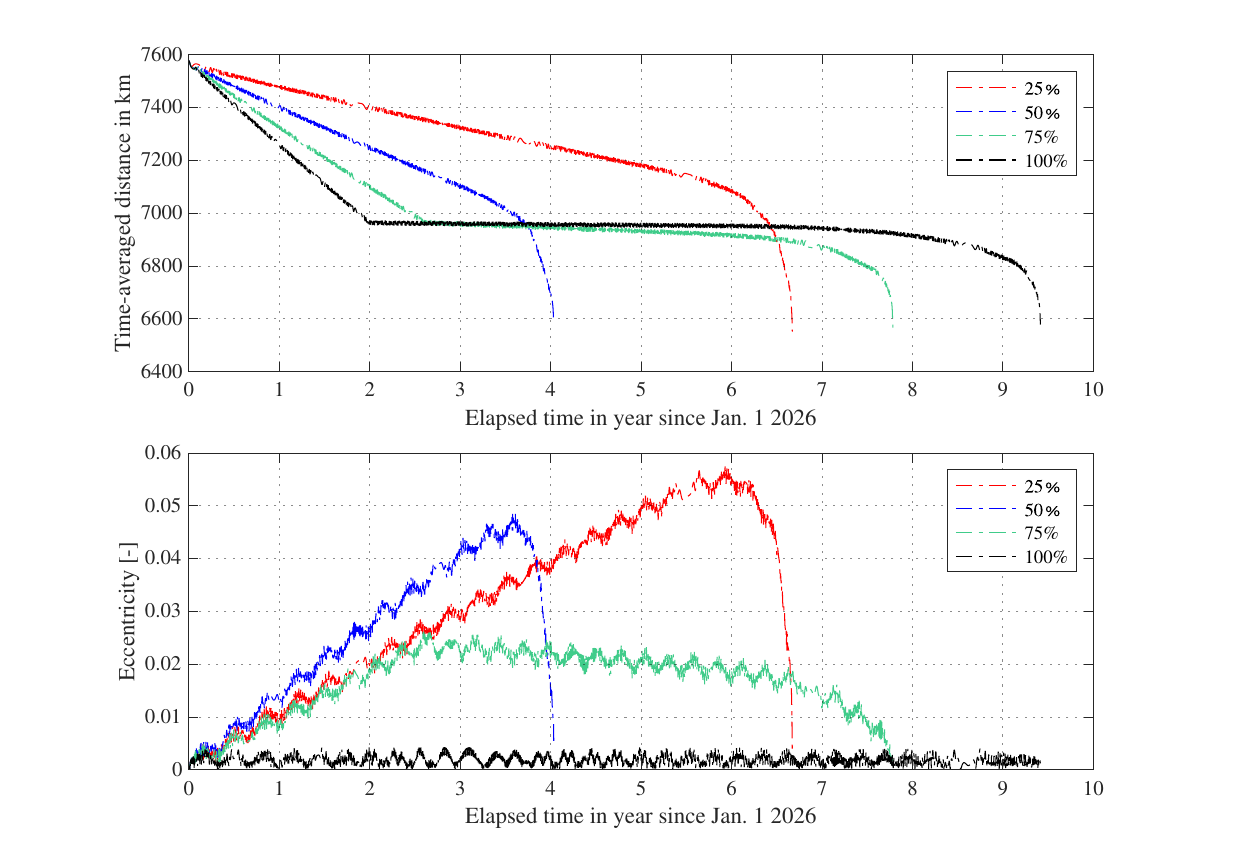}
    \caption{The change of altitude and eccentricity. }
    \label{fig:The change of altitude and eccentricity}
\end{figure}

Figure~\ref{fig:The change of altitude and eccentricity} demonstrates the altitude profile as a function of time. The variable $\bm{\bar{r}}$ is the time-averaged altitude and is used for the representative altitude for the risk assessment, because the semi-major axis oscillates during one orbital period. It is often used as a parameter for the radius of the orbit. From Kepler's equation, the orbit equation for the ellipse in terms of the eccentric anomaly is written as follows.
\begin{equation}\label{r}
    r = {a}\left(1-e\cos E\right)
\end{equation}
The time-averaged radius of an elliptical orbit $\bar{r}$ is obtained as
\begin{equation}\label{rbar}
    {\bar{r}} = \frac{1}{T}{\int_{0}^{T}rdt}\\
    = a\left(1+\frac{e^2}{2}\right)
\end{equation}
where time $t$ can be described with the orbital period $T$ and the eccentric anomaly $E$. Upper part of Fig.~\ref{fig:The change of altitude and eccentricity} shows that the deorbit behavior differs depending on the thrusting range. In the 100$\%$ and 75$\%$ scenarios, the decreasing rate of altitude significantly declines at a certain points, and after that it took 6 to 7 years to re-entry. This means that the laser satellite uses up the available thrusting time within the first few years, and after that orbital decay relies mainly on the perturbation force such as atmospheric drag. On the other hand, in the 50$\%$ and 25$\%$ scenarios, there was no significant decrease of the decreasing rate of altitude. Namely, the laser satellite can use its thrust without using up all of available thrusting time by the end of the mission in the 50$\%$ and 25$\%$ scenarios. 

Lower part of Fig.\ref{fig:The change of altitude and eccentricity} which shows the history of the eccentricity change for each scenario. From the figure it is observed that there is a tendency that the narrower the thrusting range, the larger the eccentricity reaches. When it comes to the 100$\%$ scenario, the eccentricity is almost constant. 
In 25\%, 50\% and 75\% scenarios, it is observed that the eccentricity declines just before the re-entry. This is because the difference between the perigee and apogee altitudes is much smaller near re-entry.

From these results, the relationship between the laser thrusting time and the time to reentry is examined. If the thrusting range around the apogee is narrow as in the case of 25$\%$ scenario, the apogee altitude is maintained, and only the perigee altitude is reduced preferentially. In consequence, the deorbit can be achieved while saving the laser thrusting time. On the other hand, if the thrusting range is somewhat wide, such as 50$\%$ scenario, the target orbit can be decayed at a faster rate. However, if the range is too wide, the laser thrusting time may be used up in the first few years. As shown in the examples of 75$\%$ and 100$\%$ scenarios, it will take a long time to deorbit if relying only on the perturbation force after the laser thrusting time used up. 

Table~\ref{tab:Actual_thrust_time} summarizes the total laser thrusting time and the time to reentry. In the 25$\%$ scenario, the total thrusting time is 1.7 years while the time to deorbit is 6.7 years. In contrast, in the 50$\%$ scenario, the thrusting time is almost 2 years and the time to deorbit is 4.1 years. This result shows that the 50$\%$ scenario is the best in terms of the time to reentry. In other words, if the thrusting range is set to wider than 50$\%$, the laser thrusting time will run out faster, and if the thrusting range is set to narrower than 50$\%$, the laser thrusting time will be remained. The longer the mission time, the higher the probability of satellite malfunction. Therefore, it is concluded that 50$\%$ scenario is the  best for the mission.

\begin{table}[tb]
    \centering
    \caption{Actual thrust time and the time to re-entry.
    }\scalebox{0.8}{
\begin{tabular}{ccccc} \hline \hline
Thrust range [$\%$] & Total thrust time [year] & Time to re-entry [year]\\ \hline
25 &1.7  &6.7\\
50 &2.0  &4.1\\
75 &2.0  &7.9\\
100 &2.0  &8.9\\ \hline \hline
\end{tabular}}
    \label{tab:Actual_thrust_time}
\end{table}


\subsection{Risk assessment}
In this section, the risk assessment for each possible deorbit scenario is conducted. 
In the analysis using MASTER--2009, the nominal outer space activities (Business-as-usual) is used and the other conditions and assumptions to obtain debris flux are summarised in Table~\ref{tab:risk_condition}. 
Tables~\ref{tab:res_100}-\ref{tab:res_25} show the results of collision probability along with the expected number of fragments, and Table~\ref{tab:risk_results} compares the results for the entire mission.
The variables $N$ and $ N_{\rm catas.}$ in Tables~\ref{tab:res_100}-\ref{tab:res_25} are the mean number of collision and the mean number of catastrophic collision during the time spent in each altitude zone, respectively.  
Also, the variables $P(X\geq1)$ and $P_{\mathrm{catas.}}(X\geq1)$ 
are the probability of at least one collision or catastrophic collision during the time spent in each altitude zone, respectively and the expected number of fragments is the fragments can be generated during the time spent in each altitude zone.
Note that as shown in Table~\ref{tab:risk_condition}, the characteristic length $L_c$ is 1 mm, which means that the fragments larger than 1 mm is considered.

The probability of at least one collision ($P(X\geq1)$) is the minimum in the 50\% scenario,  increases in the order of 75\%, 100\%, 25\% scenario. This order is similar to the order of the time to deorbit.
This characteristic  can be explained by Eq.~\eqref{N_def}, which indicates that the mean number of collisions is proportional to time to deorbit.
Although a similar trend is found in the probability of at least one catastrophic collision $P_{\rm cata}$, this trend is not as pronounced as the one found in $P(X>1)$.
Since the mission considered in this paper is to deorbit the target satellite, the mission success rate should be decided on the basis of the probability of catastrophic collision. 
According to the ISO's space debris mitigation guidelines, the probability of accidental collision is limited up to 0.1$\%$. As seen in Table~\ref{tab:risk_results}, even the highest $P_{\rm cata}$ is smaller than this criteria. Hence, it can be said that the target satellites could be safely deorbited without any major problems such as destruction including catastrophic fragmentation events. This demonstrates the possibility of a satellite removal mission with low thrust.

\begin{table}[tb]
\centering
\caption{Simulation parameters in MASTER--2009.}
\begin{tabular}{cc}\hline \hline
Simulation period & 2026/01/01 to re-entry \\
Satellite mass [kg] & 150 \\
Debris size threshold [m]  & 1.0$\times 10^{-3}$ -- 10  \\
Fragments size threshold (Characteristic length) $L_c$[mm] &  $\geq$1  \\
\hline \hline
\end{tabular}
\label{tab:risk_condition}
\end{table}

\begin{table}[tb]
\centering
\caption{Risk assessment results of 100$\%$ scenario.}\scalebox{0.7}{
\begin{tabular}{ccccccc}
\hline \hline
Representative altitude [km]& Duration of stay [day] &  $N$ & $N_{\mathrm{catas.}}$  & $P(X\geq1)$  & $P_{\mathrm{catas.}}(X\geq1)$ & The expected number of fragments \\ \hline 
1150  & 103.5    & 3.16E-03 & 4.07E-06 & 3.16E-03 & 4.07E-06 & 10.13117825 \\
1050  & 118.75   & 4.51E-03 & 7.98E-06 & 4.50E-03 & 7.98E-06 & 14.71061293 \\
950   & 117.5    & 6.42E-03 & 2.53E-05 & 6.40E-03 & 2.53E-05 & 91.88845826 \\
850   & 122.5    & 7.86E-03 & 3.45E-05 & 7.83E-03 & 3.45E-05 & 85.54066136 \\
750   & 122.5    & 7.68E-03 & 3.66E-05 & 7.65E-03 & 3.66E-05 & 94.35876823 \\
650   & 126.5    & 5.91E-03 & 2.13E-05 & 5.89E-03 & 2.13E-05 & 64.54914671 \\
550   & 2238.75  & 5.41E-02 & 2.39E-04 & 5.27E-02 & 2.39E-04 & 872.163191  \\
450   & 243.75   & 2.78E-03 & 1.09E-05 & 2.78E-03 & 1.09E-05 & 47.41441317 \\
350   & 45       & 1.36E-04 & 5.17E-07 & 1.36E-04 & 5.17E-07 & 1.44586067  \\
250   & 8.75     & 4.37E-06 & 1.65E-08 & 4.37E-06 & 1.65E-08 & 0.087041439 \\ \hline
TOTAL & 3247.5   & 9.26E-02 & 3.80E-04 & 8.84E-02 & 3.80E-04 & 1282.289332 \\ \hline \hline
\hline
\end{tabular}}
\label{tab:res_100}
\end{table}

\begin{table}[tb]
\centering
\caption{Risk assessment results of 50$\%$ scenario.} \scalebox{0.7}{
\begin{tabular}{ccccccc} \hline \hline
Representative altitude [km]& Duration of stay [day] &  $N$ & $N_{\mathrm{catas.}}$  & $P(X\geq1)$  & $P_{\mathrm{catas.}}(X\geq1)$ & The expected number of fragments \\ \hline 
1150  & 188.75   & 5.69E-03 & 8.46E-06 & 5.67E-03 & 8.46E-06 & 22.06207508 \\
1050  & 226.25   & 9.95E-03 & 2.49E-05 & 9.90E-03 & 2.49E-05 & 72.51149657 \\
950   & 238.75   & 1.41E-02 & 4.18E-05 & 1.40E-02 & 4.18E-05 & 118.6434662 \\
850   & 240      & 1.53E-02 & 4.77E-05 & 1.52E-02 & 4.77E-05 & 121.992824  \\
750   & 245      & 1.35E-02 & 5.26E-05 & 1.34E-02 & 5.26E-05 & 130.4921677 \\
650   & 187.5    & 7.85E-03 & 3.23E-05 & 7.82E-03 & 3.23E-05 & 90.18908973 \\
550   & 68.75    & 2.00E-03 & 9.08E-06 & 2.00E-03 & 9.08E-06 & 27.57261229 \\
450   & 37.5     & 5.71E-04 & 2.71E-06 & 5.71E-04 & 2.71E-06 & 10.28657394 \\
350   & 27.5     & 1.48E-05 & 4.82E-07 & 1.48E-05 & 4.82E-07 & 1.867874963 \\
250   & 10       & 1.21E-05 & 2.90E-08 & 1.21E-05 & 2.90E-08 & 0.075360131 \\ \hline
TOTAL & 1470     & 6.90E-02 & 2.20E-04 & 6.67E-02 & 2.20E-04 & 595.6935407 \\ \hline \hline
\end{tabular}}
\label{tab:res_50}
\end{table}

\begin{table}[tb]
\centering
\caption{Risk assessment results of 75$\%$ scenario.}\scalebox{0.7}{
\begin{tabular}{ccccccc} \hline\hline
Representative altitude [km]& Duration of stay [day] &  $N$ & $N_{\mathrm{catas.}}$  & $P(X\geq1)$  & $P_{\mathrm{catas.}}(X\geq1)$ & The expected number of fragments \\ \hline 
1150  & 130      & 4.07E-03 & 5.80E-06 & 4.07E-03 & 5.80E-06 & 15.43495741 \\
1050  & 155      & 6.73E-03 & 1.39E-05 & 6.71E-03 & 1.39E-05 & 34.79796901 \\
950   & 156.25   & 8.48E-03 & 2.92E-05 & 8.45E-03 & 2.92E-05 & 104.182181  \\
850   & 162.5    & 1.08E-02 & 3.66E-05 & 1.07E-02 & 3.66E-05 & 92.46511698 \\
750   & 162.5    & 8.96E-03 & 3.64E-05 & 8.92E-03 & 3.64E-05 & 87.3052741  \\
650   & 167.5    & 7.23E-03 & 3.08E-05 & 7.20E-03 & 3.08E-05 & 94.96667183 \\
550   & 1597.5   & 3.92E-02 & 1.91E-04 & 3.85E-02 & 1.91E-04 & 738.1165488 \\
450   & 241.25   & 2.41E-03 & 1.06E-05 & 2.41E-03 & 1.06E-05 & 53.85361527 \\
350   & 60       & 1.81E-04 & 6.49E-07 & 1.81E-04 & 6.49E-07 & 2.707332387 \\
250   & 8.75     & 4.48E-06 & 1.57E-08 & 4.48E-06 & 1.57E-08 & 0.07755249  \\ \hline
TOTAL & 2841.25  & 8.81E-02 & 3.55E-04 & 8.43E-02 & 3.55E-04 & 1223.907219 \\ \hline \hline
\end{tabular}}
\label{tab:res_75}
\end{table}

\begin{table}[tb]
\centering
\caption{Risk assessment results of 25$\%$ scenario.}
\scalebox{0.7}{
\begin{tabular}{ccccccc} \hline \hline
Representative altitude [km]& Duration of stay [day] &  $N$ & $N_{\mathrm{catas.}}$  & $P(X\geq1)$  & $P_{\mathrm{catas.}}(X\geq1)$ & The expected number of fragments\\ \hline 
1150  & 372.5    & 1.15E-02 & 1.50E-05 & 1.15E-02 & 1.50E-05 & 37.6192412  \\
1050  & 461.25   & 1.99E-02 & 5.20E-05 & 1.97E-02 & 5.20E-05 & 151.5199467 \\
950   & 493.75   & 2.77E-02 & 8.74E-05 & 2.73E-02 & 8.74E-05 & 237.349526  \\   
850   & 507.5    & 2.58E-02 & 8.31E-05 & 2.54E-02 & 8.31E-05 & 207.4551961 \\
750   & 365      & 1.82E-02 & 6.55E-05 & 1.80E-02 & 6.55E-05 & 175.8183456 \\
650   & 127.5    & 5.74E-03 & 2.20E-05 & 5.73E-03 & 2.20E-05 & 62.11856449 \\
550   & 56.25    & 1.67E-03 & 7.41E-06 & 1.66E-03 & 7.41E-06 & 22.28920811 \\
450   & 26.25    & 3.22E-04 & 1.74E-06 & 3.22E-04 & 1.74E-06 & 7.47004171  \\
350   & 20       & 8.30E-05 & 2.82E-07 & 8.30E-05 & 2.82E-07 & 1.343879557 \\
250   & 5        & 3.66E-06 & 1.11E-08 & 3.66E-06 & 1.11E-08 & 0.045478762 \\  \hline
TOTAL & 2435     & 1.11E-01 & 3.35E-04 & 1.05E-01 & 3.34E-04 & 903.0294282 \\ \hline \hline
\end{tabular}}
\label{tab:res_25}
\end{table}

\begin{table}[tb]
    \centering
    \caption{Comparison of risk assessment results among 4 scenarios(overall mission)}    
\begin{tabular}{ccccc} \hline \hline
scenario & Time to deorbit [year] & $P(X\geq 1)$ & $P_{\rm cata.}$ & The expected number of fragments\\ \hline
100$\%$ &8.9  &8.84  &0.038  & 1282.29 \\
75$\%$ &7.8  &8.43  &0.035  & 1233.91\\
50$\%$ &4.0  &6.67  &0.022  & 595.69 \\
25$\%$ &6.7  &11.08  &0.034  & 903.03 \\ \hline \hline
\end{tabular}
    \label{tab:risk_results}
\end{table}

\begin{table}[tb]
\centering
\caption{Comparison of risk assessment results among 4 scenarios(per year).} 
\begin{tabular}{ccc}\hline \hline
scenarios & $N$ per year                   & The expected number of fragments per year \\ \hline
25 $\%$   & 1.67E-02                     & 135.45       \\
50 $\%$   & 1.71E-02                     & 148.01       \\
75 $\%$   & 1.13E-02                     & 157.34       \\
100 $\%$  & 1.04E-02 & 144.22      \\ \hline \hline
\end{tabular}
\label{tab:results_per_year}
\end{table}

A study that analyzed the same orbit and satellite size as this study~\cite{radtke2017interactions} showed that if the active disposal is completed successfully, $P(X\geq1)$ is calculated as 0.12 $\%$, which is far below the results in Table \ref{tab:risk_results}. 
This can be explained by the prompt deorbit and the size of debris to be considered. 
For instance, OneWeb satellites under nominal conditions can complete their active disposal within 2 years by using their own ion thrusters, which can generate stronger thrust than that used in this study.
Since the time to deorbit strongly affects the collision risk, this could explain this difference. 
In addition, \cite{radtke2017interactions} considers debris larger than 1 cm, whereas this paper considers debris larger than 0.1 cm, which also could contribute to the lower risk.   

\begin{figure}[tb]
    \centering
    \includegraphics[scale = 0.7]{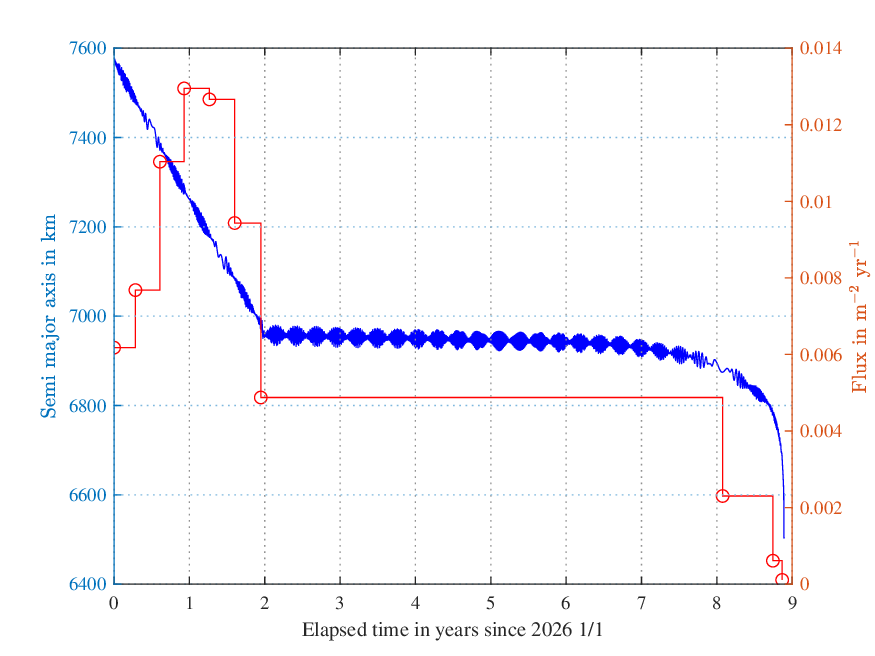}
    \caption{Orbital trajectories along with the flux value at each altitude zone. (100$\%$ scenario)}
    \label{fig:fx_circ}
\end{figure}

\begin{figure}[tb]
    \centering
    \includegraphics[scale = 0.7]{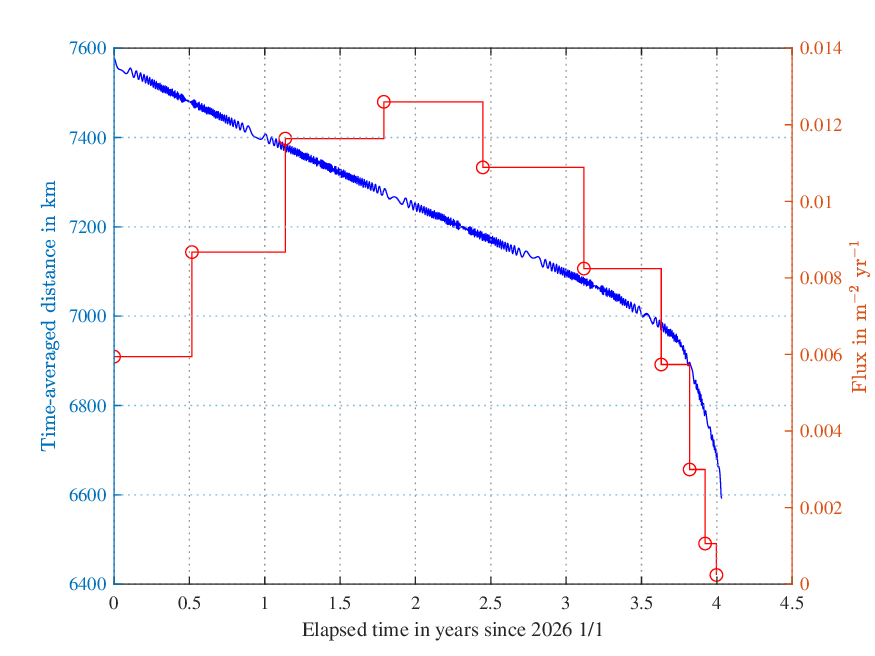}
    \caption{Orbital trajectories along with the flux value at each altitude zone. (50$\%$ scenario)}
    \label{fig:fx_ellip}
\end{figure}

\begin{figure}[tb]
    \centering
    \includegraphics[scale = 0.7]{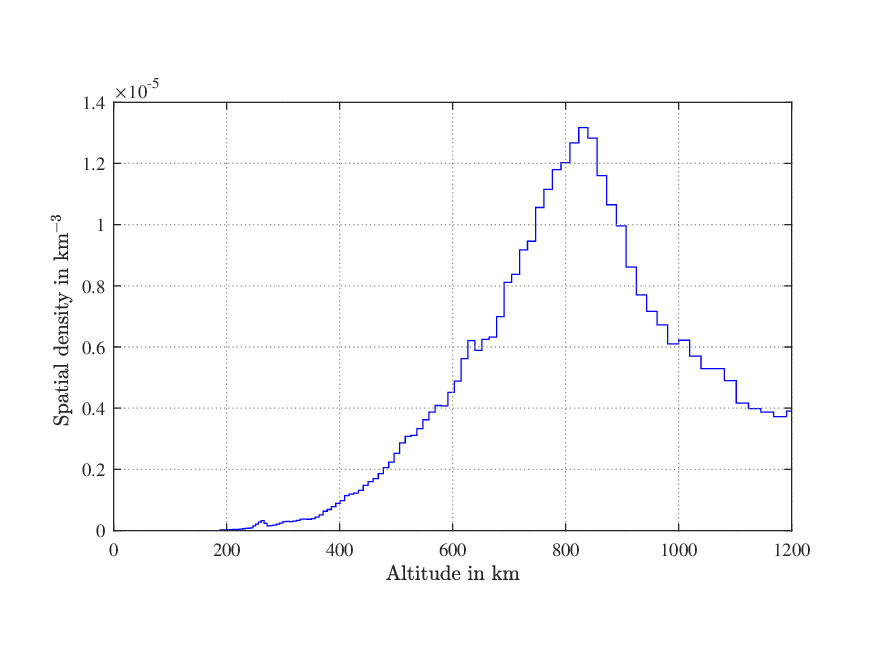}
    \caption{Spatial density as a function of altitude.}
    \label{fig:spatial_density}
\end{figure}

Figures~\ref{fig:fx_circ} and~\ref{fig:fx_ellip} represent the orbital trajectories along with the flux values at each altitude zone. In both scenarios, high flux values at around 800 km are observed. 
Figure~\ref{fig:spatial_density} shows the spatial density at different altitudes. One can find that the spatial density is the highest at around the altitude of 800 km, which corresponds to the high flux altitude area observed in Figs.~\ref{fig:fx_circ} and~\ref{fig:fx_ellip}.

In~\ref{subsection:O-P}, it can be seen that the altitude change rate is almost proportional to the thrusting range in the region where laser thrusting is valid. 
The sooner the satellites exit this altitude zone, the less risk it will be. 
However, the results in Table~\ref{tab:risk_results} indicate that the 50$\%$ scenario has the lowest risk. 
Even though the satellite exits this altitude band the fastest in the 100$\%$ thrust scenario,  it takes 6 more years to decay from 600 km because of the low level of atmospheric drag at 600 km. The trade-off problem thus arises between the low thrust time constraint and time to deorbit. 

The collision risk on a yearly basis is summarized in Table~\ref{tab:results_per_year}. Interestingly enough, the results show that the values in the 50 $\%$ scenario are the highest, and the values in the 100 $\%$ scenario are the lowest, which is contrary to the results shown in Table~\ref{tab:risk_results}. Possible reasons for this difference could be the time to deorbit 
versus the time spent in the altitude of around 800 km, where the higher spatial density is observed in Fig.~\ref{fig:spatial_density}. 
As seen in Fig.~\ref{fig:The change of altitude and eccentricity}, the altitude change rate is almost proportional to the thrusting range. 
For example, in the 100$\%$ scenario, the time to deorbit is long, and the time spent at the altitude with high flux is relatively short, and the risk per year is small, but the overall risk is high.
In the 50$\%$ scenario, the duration stayed in that high-flux area is longer and the time to deorbit is shorter than the 100$\%$ scenario. Hence, the unit-year risk is found to be larger, and the overall risk is found to be less. 

\begin{figure}[tb]
    \centering
    \includegraphics[scale = 0.7]{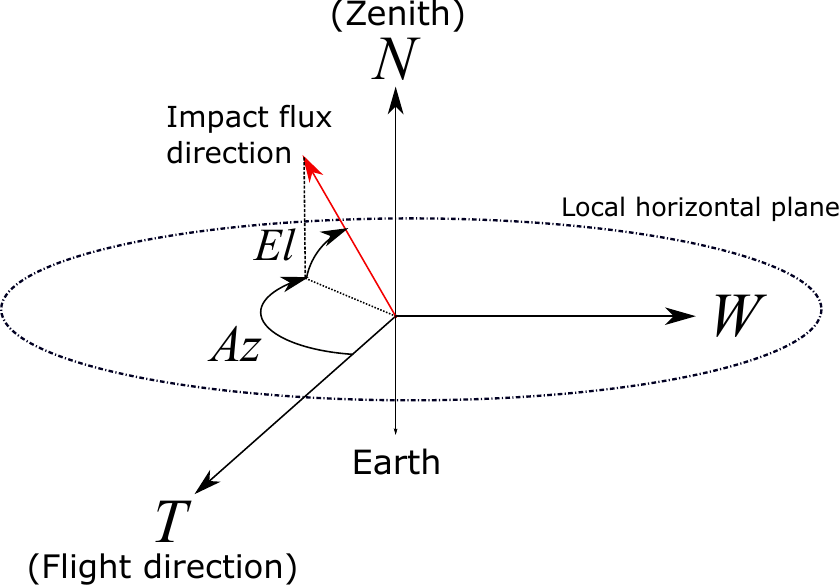}
    \caption{Definition of impact azimuth ($Az$) and elevation ($El$) angle.}
    \label{fig:angle_def}
\end{figure}

\begin{figure}[tb]
    \centering
    \includegraphics[scale = 0.7]{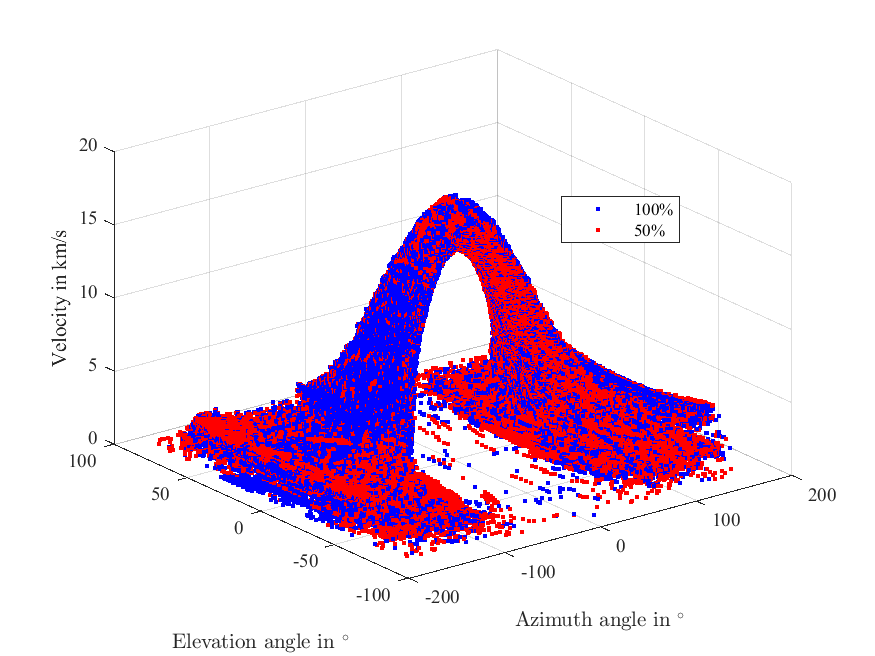}
    \caption{Distribution of impact angles (Azimuth and elevation) and relative velocity for 50$\%$ and 100$\%$ scenarios.}
    \label{fig:3D}
\end{figure}

\begin{figure}[tb]
    \centering
    \includegraphics[scale = 0.5]{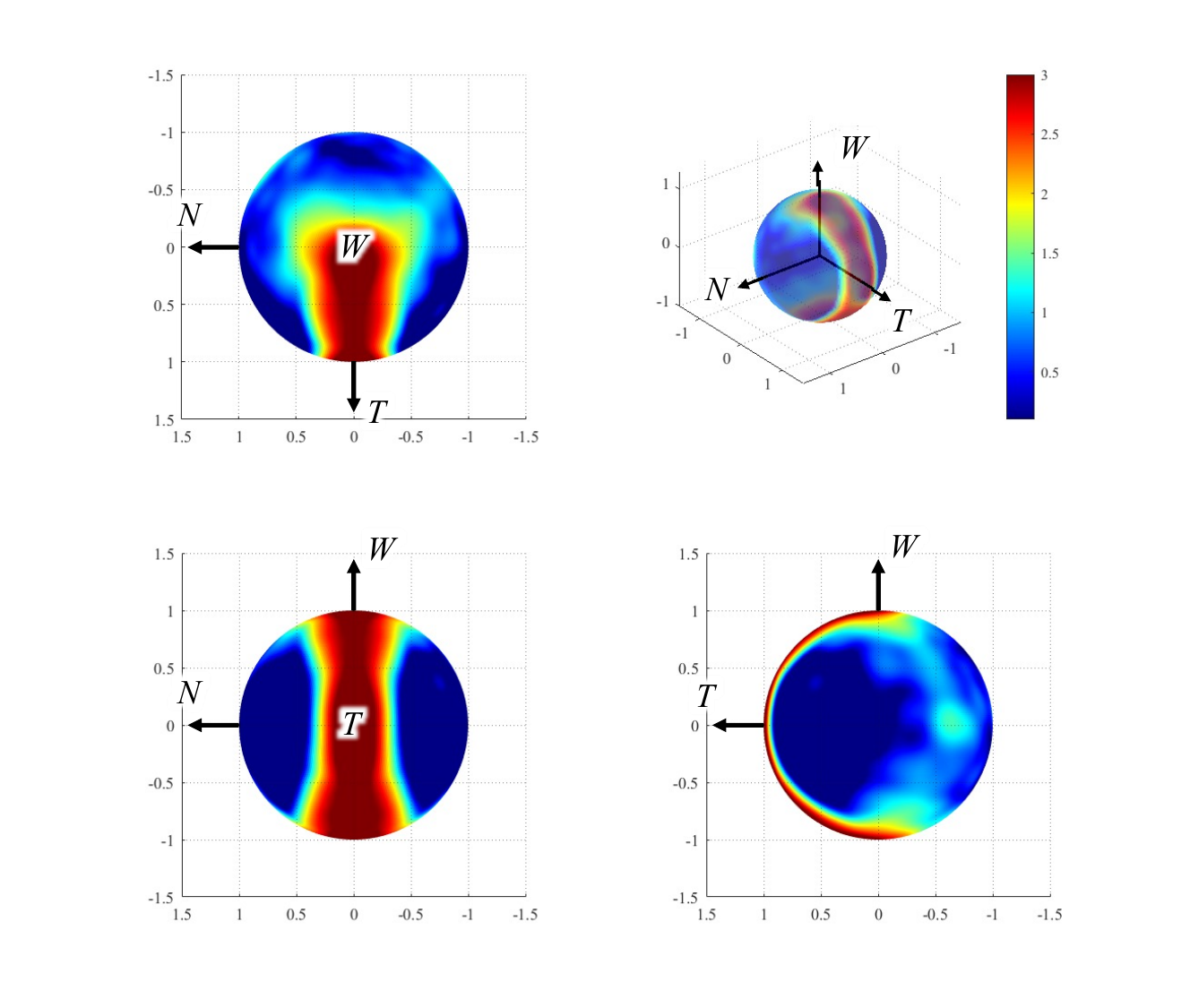} 
    \caption{Spherical plot of impact angle distribution for the 50$\%$ scenario}
    \label{fig:heatmap}
\end{figure}

Figure~\ref{fig:3D} shows the relationship between the impact angles (azimuth and elevation angles) and the relative velocity for the 100$\%$ and 50$\%$ scenarios. The definition of azimuth and elevation angles is illustrated in Fig.~\ref{fig:angle_def}, where for example a debris fragment approaching the target satellite from the positive in-track direction would have an azimuth and elevation of zero. 
As for these impact angles, no significant difference is observed. 
This is because the eccentricity reaches 0.06 even in the maximum thrust case, which is small enough to be considered as a circular orbit. 
In other words, there is little difference in the collision characteristics of the 4 scenarios due to the difference in orbit shape.  

The eccentric decay has two additional potential risks.
The one is that in an elliptical orbit, velocity is increased when passing through the perigee. The other one is that the elliptical orbit of the target satellite crosses the circular orbit that most debris have. These risks can contribute to the high risk of collision. 
However, from these results, this concern can be ruled out. Figure~\ref{fig:heatmap}
represents the spherical distribution of impact angle for the 50$\%$ scenario. 
High collision frequency from the direction of travel, above and below is observed. 
For the same reason, only the case of 50$\%$ scenario is presented here, as there was no significant difference in the distribution of impact angles among 4 scenarios.

In terms of possible environmental effects, the expected number of fragments during the decay are also compared in Table~\ref{tab:risk_results}. 
Along with the collision probabilities, the number of fragments generated during decay shows a similar trend. This is due to the definition of the expected number of fragments, which is directly computed from its collision frequency.
Even in the case of the lowest number of fragments, almost 600 fragments can be generated during its decay. 
As shown in Eq.~\eqref{Exp_def}, the expected number of fragments $\mathbb{E}_{\rm tot}$ takes into account all possible events, this value means that the number of fragments that can be generated under the worst case conditions. Thus, the value of $\mathbb{E}_{\rm tot}$ is relatively large while the value of $P(X\geq 1)$ is small.   
From the discussion above, the satellites should be designed to deorbit as soon as possible in terms of the orbital environment preservation. 

\clearpage
\section{Conclusions}
This paper assessed the collision risk for deorbit missions with low-thrust deceleration and revealed the relationship between the time to deorbit and collision risk. 
As a preliminary step for the risk assessment, the relationship between the range of low thrust and the time to deorbit is obtained by numerical simulations. It is revealed that the 50$\%$ scenario is the most effective to deorbit the target in terms of deorbiting time in this study. 
The probability for at least one collision mainly depends on the time to deorbit, and collision risk increases as the time to re-entry becomes longer. The expected number of fragments is investigated and was found to increase for longer mission duration. This is because the mean number of collisions is proportional to the duration. 
No significant difference in impact angles and relative velocity for each mission was observed, due to the low eccentricity. 
As for catastrophic collision, it 
was demonstrated by our results that the likelihood of catastrophic collision was low enough to meet the ISO's space mitigation guidelines.  
This fact also revealed that the EOL mission with low thrust could be carried out in a safe manner.

\clearpage
\bibliography{tether}

@article{virgili2016risk,
  title={Risk to space sustainability from large constellations of satellites},
  author={Virgili, B Bastida and Dolado, JC and Lewis, HG and Radtke, J and Krag, H and Revelin, B and Cazaux, C and Colombo, CAMILLA and Crowther, R and Metz, M},
  journal={Acta Astronautica},
  volume={126},
  pages={154--162},
  year={2016},
  publisher={Elsevier}
}

@book{Klinkrand,
  author    = {Heiner Klinkrad}, 
  title     = {Space Debris- Models and Risk Analysis},
  publisher = {Springer-Verlag Berlin Heidelberg},
  year      = 2006,
  volume    = 4,
  series    = 10,
  address   = {The address},
  edition   = 1,
  month     = 7,
  note      = {},
  isbn      = {978-3-642-42623-0}
}

@article{radtke2017interactions,
  title={Interactions of the space debris environment with mega constellations—Using the example of the OneWeb constellation},
  author={Radtke, Jonas and Kebschull, Christopher and Stoll, Enrico},
  journal={Acta Astronautica},
  volume={131},
  pages={55--68},
  year={2017},
  publisher={Elsevier}
}

@article{johnson2001nasa,
  title={NASA's new breakup model of EVOLVE 4.0},
  author={Johnson, Nicholas L and Krisko, PH and Liou, J-C and Anz-Meador, PD},
  journal={Advances in Space Research},
  volume={28},
  number={9},
  pages={1377--1384},
  year={2001},
  publisher={Elsevier}
}

@inproceedings{alfano2020leo,
  title={LEO constellation encounter and collision rate estimation: an update},
  author={Alfano, S and Oltrogge, DL and Shepperd, R},
  booktitle={2nd IAA Conference on Space Situational Awareness (ICSSA), Washington DC, IAA-ICSSA-20-0021},
  volume={15},
  year={2020},
  organization={ICSSA}
}

@inproceedings{oltrogge2016collision,
  title={Collision risk in low earth orbit},
  author={Oltrogge, DL and Alfano, S},
  booktitle={2016 International Astronautical Congress, Guadalajara, Mexico},
  volume={23},
  year={2016}
}

@article{le2018space,
  title={Space debris collision probability analysis for proposed global broadband constellations},
  author={Le May, S and Gehly, S and Carter, BA and Flegel, S},
  journal={Acta Astronautica},
  volume={151},
  pages={445--455},
  year={2018},
  publisher={Elsevier}
}

@article{KAWAMOTO2020178,
title = {Impact on collision probability by post mission disposal and active debris removal},
journal = {Journal of Space Safety Engineering},
volume = {7},
number = {3},
pages = {178-191},
year = {2020},
note = {Space Debris: The State of Art},
issn = {2468-8967},
doi = {https://doi.org/10.1016/j.jsse.2020.07.012},
url = {https://www.sciencedirect.com/science/article/pii/S246889672030077X},
author = {Satomi Kawamoto and Nobuaki Nagaoka and Tsuyoshi Sato and Toshiya Hanada}
}

@article{sims1999preliminary,
  title={Preliminary design of low-thrust interplanetary missions},
  author={Sims, J and Flanagan, S},
  year={1999}
}

@article{baldwin2012optimal,
  title={Optimal deorbit guidance},
  author={Baldwin, Morgan C and Lu, Ping},
  journal={Journal of guidance, control, and dynamics},
  volume={35},
  number={1},
  pages={93--103},
  year={2012}
}

@article{2004orbital,
  title={Orbital debris quarterly news},
  author={J.-C. Liou, Sara Portman},
  year={2004}
}

@article{tsuno2020impulse,
  title={Impulse measurement of laser induced ablation in a vacuum},
  author={Tsuno, Katsuhiko and Wada, Satoshi and Ogawa, Takayo and Ebisuzaki, Toshikazu and Fukushima, Tadanori and Hirata, Daisuke and Yamada, Jun and Itaya, Yuki},
  journal={Optics Express},
  volume={28},
  number={18},
  pages={25723--25729},
  year={2020},
  publisher={Optical Society of America}
}

@article{RHATIGAN2020340,
title = {Drag-enhancing deorbit devices for spacecraft self-disposal: A review of progress and opportunities},
journal = {Journal of Space Safety Engineering},
volume = {7},
number = {3},
pages = {340-344},
year = {2020},
note = {Space Debris: The State of Art},
issn = {2468-8967},
doi = {https://doi.org/10.1016/j.jsse.2020.07.026},
url = {https://www.sciencedirect.com/science/article/pii/S2468896720300896},
author = {Jennifer L. Rhatigan and Wenschel Lan}
}

@article{SATO2020813,
title = {Performance of EDT system for deorbit devices using new materials},
journal = {Acta Astronautica},
volume = {177},
pages = {813-820},
year = {2020},
issn = {0094-5765},
doi = {https://doi.org/10.1016/j.actaastro.2020.01.029},
url = {https://www.sciencedirect.com/science/article/pii/S0094576520300400},
author = {Tsuyoshi Sato and Satomi Kawamoto and Yasushi Ohkawa and Takeo Watanabe and Koh Kamachi and Hiroshi Okubo}
}

@article{phipps1996orion,
  title={ORION: Clearing near-Earth space debris using a 20-kW, 530-nm, Earth-based, repetitively pulsed laser},
  author={Phipps, CR and Albrecht, G and Friedman, H and Gavel, D and George, EV and Murray, J and Ho, C and Priedhorsky, W and Michaelis, MM and Reilly, JP},
  journal={Laser and Particle Beams},
  volume={14},
  number={1},
  pages={1--44},
  year={1996},
  publisher={Cambridge University Press}
}

@inproceedings{bischof2003roger,
  title={ROGER-Robotic geostationary orbit restorer},
  author={Bischof, Bernd},
  booktitle={54th International Astronautical Congress of the International Astronautical Federation, the International Academy of Astronautics, and the International Institute of Space Law},
  pages={IAA--5},
  year={2003}
}

@article{anz2018orbital,
  title={Orbital debris quarterly news},
  author={Anz-Meador, Phillip D},
  year={2018}
}

@article{lemmens2019esa,
  title={{ESA} annual space environment report},
  author={Lemmens, Stijn and Letizia, Francesca},
  journal={ESA Space Debris Office, Darmstadt, Germany, Tech. Rep. GEN-DB-LOG-00271-OPS-SD},
  year={2019}
}

\end{document}